\documentclass[10pt, conference, compsocconf]{IEEEtran}
\ifCLASSINFOpdf
\else
\fi

\usepackage{microtype}      
\usepackage{mathrsfs}  
\usepackage{amssymb,amsmath}
\usepackage{graphicx}
\usepackage{hyperref}
\usepackage{subcaption}
\usepackage{wrapfig}
\usepackage{picinpar}
\usepackage{cutwin}
\usepackage{graphicx,multicol}
\usepackage{algorithm}  
\usepackage{algorithmic}
\usepackage{amsfonts}
\usepackage{amsfonts,amssymb}
\usepackage{url}

\begin{document}
%
\title{Corticospinal Tract (CST) reconstruction based on fiber orientation distributions (FODs) tractography}


\author{\IEEEauthorblockN{Authors Name/s per 1st Affiliation (Author)}
\IEEEauthorblockA{line 1 (of Affiliation): dept. name of organization\\
line 2: name of organization, acronyms acceptable\\
line 3: City, Country\\
line 4: Email: name@xyz.com}
}

\author{\IEEEauthorblockN{Youshan Zhang}
\IEEEauthorblockA{Computer Science and Engineering Department\\
Lehigh University\\
Bethlehem, PA\\
yoz217@lehigh.edu}
}


%


\maketitle

\begin{abstract}
The Corticospinal Tract (CST) is a  part  of  pyramidal tract (PT) and it can innervate the voluntary movement of skeletal muscle through spinal interneurons (the 4th layer of the Rexed gray board layers), and anterior horn motorneurons (which control trunk and proximal limb muscles). Spinal  cord  injury  (SCI)  is  a  highly disabling disease often caused by traffic accidents. The recovery of CST and the functional reconstruction of spinal anterior horn motor neurons play an essential role in the treatment of  SCI.  However,  the  localization  and  reconstruction of CST are still challenging issues, the accuracy of the geometric  reconstruction  can  directly  affect  the  results  of the surgery. The main contribution of this paper is the reconstruction of the CST based on the fiber orientation distributions (FODs) tractography. Differing from tensor-based tractography in which the primary direction is a determined orientation, the direction of FODs tractography is determined by the probability. The spherical harmonics (SPHARM) can be used to approximate the efficiency of FODs tractography. We manually delineate the three ROIs (the posterior limb of the internal capsule, the cerebral peduncle, and the anterior pontine area) by the ITK-SNAP software, and use the pipeline software to reconstruct both the left and right sides of the CST fibers. Our results demonstrate that FOD-based tractography can show more and correct anatomical CST fiber bundles. 

\end{abstract}


%
\IEEEpeerreviewmaketitle

\section{Introduction}
   The Corticospinal Tract (CST) is a part of pyramidal tract (PT), and many of its fibers originate from the primary motor cortex (the precentral gyrus) then terminate in the spinal cord \cite{lemon2005comparing, jang2014corticospinal}. The CST is thought to originate from the premotor area, the supplementary motor area, and the somatosensory cortex, and then terminate at the thoracic levels (20\%), lumbosacral levels (25\%) and cervical levels (55\%) \cite{lemon2008descending,masdeu2011localization}. Evidence shows that the CST serves as the major downstream motor tract in the mammalian spinal cord, and the fibers control the voluntary movement of skeletal muscle by directly or indirectly innervating the neurons in the anterior horn \cite{freund2008anti}. However, the structural and functional properties of the CST are far more complex than researchers previously thought, and we continue to have a new understanding about the CST with the advancement of technology, thus the CST is still one of the hotspots in neuroscience research \cite{bigler2017incorporating,potapov2015current,weiss2015improved}.
   About 75\%-90\% of the CST fibers descend to the lateral corticospinal tract by decussation of the pyramid, but a few fibers form the anterior corticospinal tract by not crossing the pyramid \cite{misra2010study,fisch2012neuroanatomy}. The fibers, which first passed through the medial posterior spinocerebellar tract then went into the lumbosacral spinocerebellar tract, were not yet appeared 
   \cite{andersson1995physiology,watson2012basic}. The main function of the CST is innervating the voluntary movement of skeletal muscle through spinal interneurons (the 4th layer of the Rexed gray board layers), and anterior horn motorneurons (which control trunk and proximal limb muscles), then the CST terminates in the spinal motor cells (which control the fine motor of small muscle in extremities) 
   \cite{holtz2010spinal}. Spinal cord injury (SCI) is a highly disabling disease in clinical practice. With the popularization of modern transport vehicles and the acceleration of the process of industrialization, the number of patients with SCI is increasing day by day, many people were injured in traffic accidents. If the CST is damaged, it can seriously affect the quality of life of patients and even lead to paralysis. Patients with rigid muscle, spasm, paralysis and other pathologies are usually considered to have CST injury \cite{abbott1947abdominal}. When the CST is damaged in the SCI, the motor pathways are affected, and these surviving spinal cord nerve cells, which are below the injury surface, lose neural innervation of the brain, and are unable to perform the random functional activity \cite{maier2006sprouting}.
 In addition, the repair of the CST after the SCI is still a vital research topic in neuroscience, and the recovery of limb motor function can be promoted by the CST repair or functional remodeling. Although many types of experimental research have made great progress, there has been no report of successful clinical application \cite{rowland2008current}. Therefore, the recovery of CST and the functional reconstruction of spinal anterior horn motor neurons play an essential role in the treatment of SCI \cite{tohda2011current}. However, the localization and reconstruction of CST are still challenging issues, and the accuracy of the geometric reconstruction can directly affect the results of the surgery. Furthermore, there are few methods that can be used to reconstruct the CST and the validation of clinical application is still uncertain \cite{qazi2009resolving}.  
   
 \begin{figure*}[htbp!]
\centering
\includegraphics[scale=0.9]{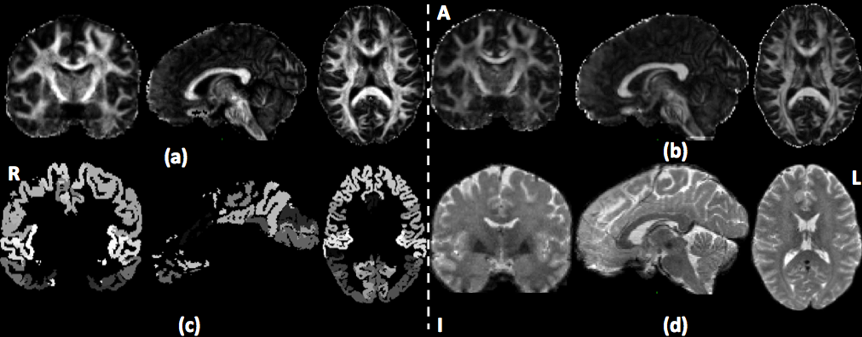}\\
\caption{ The four different images are used in the reconstruction of the CST fibers. (a): FOD images, which are the FODs tractography files, (b): FA images which can be used to draw the ROIs using ITK-SNAP, (c): The CortexLabel images that contain the labels of all the cortex; the labels of the left and right precentral gyrus are 49 and 50. (d): The mask of the tractography image. In each image, the left is the coronal slice, the middle is the sagittal slice and the right is the axial slice (A: anterior, I: inferior, R: right, L: left).   }
\label{fig:data}
\end{figure*}

\begin{figure*}[htbp!]
\centering
\includegraphics[scale=0.8]{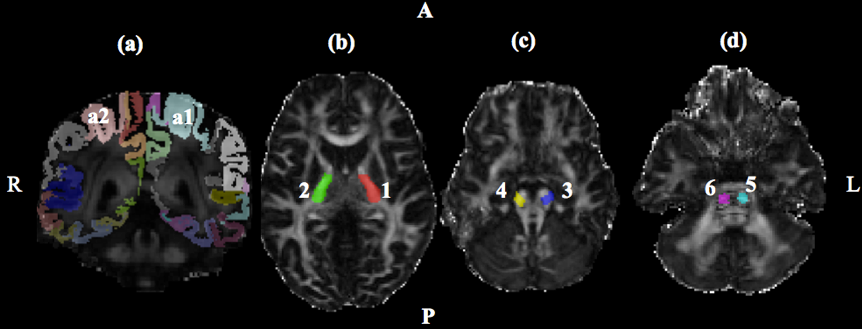}\\
\caption{The seed region and three ROIs in the reconstruction of the CST fibers. (a): the coronal slice of the seed region precentral gyrus, a1 is the left precentral gyrus and a2 is the right precentral gyrus. From (b) to (d) are the axial slice the three ROIs (the posterior limb of the internal capsule, the cerebral peduncle, and the anterior pontine area). These three ROIs are manually delineated by the ITK-SNAP. 1-6 are six labels, 1, 3 and 5 are the left ROIs, 2, 4, and 6 are the right ROIs (A: anterior, P: posterior, R: right, L: left).   }
\label{fig:mppga}
\end{figure*}
   Diffusion tensor imaging (DTI) is a non-invasive technique that can estimate the integrity of the white matter tracts by using the diffusion of water molecules \cite{mori1999three,le2001diffusion,barone2014image}. Diffusion tensor tractography (DTT), a three-dimensional visualized version of DTI, has been widely used to reconstruct the CST 
   \cite{weiss2015improved,nagae2004high,do2013injury}. However, there are several challenges in these papers. Firstly, the seed regions or the regions of interest (ROIs) are different in these papers, the models of these papers did not cover all the anatomic areas of the CST, and the results did not reflect the complexity of the CST fibers. Secondly, the validation of these models in the clinical application still not fully clarified.

The main contribution of this paper is the reconstruction of the CST based on the fiber orientation distributions (FODs) tractography
\cite{kammen2016automated,tournier2010improved}. The major difference between the FODs tractography and the tensor-based tractography is the definition of the tracts direction. Tensor-based tractography relies on the three non-negative eigenvalues and eigenvectors, the main direction is a determined orientation
\cite{tournier2011diffusion}. The FOD tractography is fiber orientation distribution (the direction is determined by the probability), the spherical harmonics (SPHARM) can be used to approximate the efficiency of FOD \cite{kammen2016automated}. In this paper, we manually delineate the ROIs by the ITK-SNAP software\cite{yushkevich2016itk}, and use the Lab of NeuroImaging (LONI) pipeline software\cite{dinov2010neuroimaging} to reconstruct both the left and right sides of the CST fibers. Our results demonstrate that FOD-based tractography can show correct anatomical CST fiber bundles. 


\begin{figure*}[htbp!]
\centering
\includegraphics[scale=1]{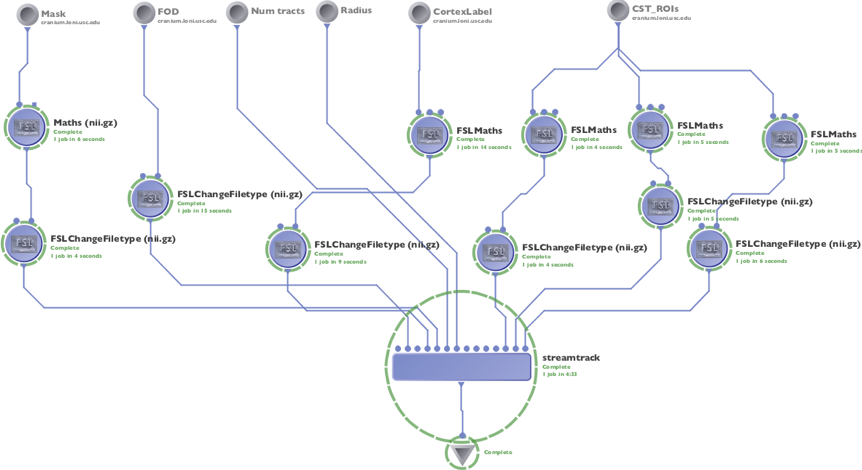}\\
\caption{The workflow of CST reconstruction method from Pipeline software, which is used to generate the reconstruction CST fibers based on the FODs tractography. The input files are the gray circle buttons (Mask, FOD, Num track, Radius, CortexLabel, and the CST\_ROIs), the output files are the gray triangular button, which is the fibers of the CST. The middle FSL Maths modules can add more images in one module, the FSLChangeFiletype module can change the file type of the image, and the streamtrack module can compute the fibers of the CST based on the FODs.  }
\label{fig:flow}
\end{figure*}

\section{Method}
\subsection{Data}
  In this paper, the data are from Human Connectome Project (HCP) \cite{fan2016mgh}, which can be found on LONI IDA (\url{https://ida.loni.usc.edu/services/NewUser.jsp}) or ConnectomeDB (\url{https://db.humanconnectome.org/app/template/Login.vm}) . Figure 1 shows the four images ( fiber orientation distributions (FOD), fractional anisotropy(FA), CortexLabel and Mask), which are used to reconstruct the CST fibers.

\subsection{ROIs} 
   The CST primarily originates from the motor cortex (the precentral gyrus), then descends to the forebrain, pons, medulla, and pyramidal decussation and terminates at thoracic levels, lumbosacral levels, and cervical levels. Therefore, the ROIs should be in these different areas, but the proper ROIs are still hard to decide. Many papers used the precentral gyrus as the starting point since precentral gyrus is the major area controls the movement of the arm, leg and the trunk \cite{cabeencorticospinal,weiss2015improved,yeo2014different}. However, the organization of the fibers can be distorted in patients with brain tumors \cite{weiss2015improved}, so more ROIs along the CST should be included. Another two ROIs (the posterior limb of the internal capsule (PLIC) and anterior pontine area (aiP)) were used in Carolin et al., 2015. The internal capsule lies in the forebrain, the pons is a part of the brainstem, both of these two regions are involved in the movement pathways, they can affect the movement of the leg, trunk, and arm. Therefore, the PLIC and aiP are two essential ROIs. Several papers showed that the cerebral peduncle (CP) of the midbrain is another important ROI of the CST reconstruction, since it also involved in innervating the movement pathways \cite{cabeencorticospinal,schaechter2009microstructural}. Although Venkateswaran and Erik, 2017 used centrum semiovale at top of lateral ventricle (CSoLV) as additional ROIs\cite{bigler2017incorporating}, the CSoLV is not significant comparing with the other three ROIs, therefore, we exclude this area in our study. In this paper, the seed region is the precentral gyrus, the ROIs are the PLIC, CP and the aiP. Figure 2 shows the seed region and ROIs in the CST reconstruction.

\subsection{Workflow of the Pipeline}
   We use Pipeline software \cite{dinov2010neuroimaging} to reconstruct the CST fibers after we choose the seed region and the ROIs. Figure 3 shows the reverent workflow of the Pipeline when we compute the fibers of the CST.

\subsection{Parameters}
   The input files Mask, FOD, and CortexLabel are the three images which are mentioned in section A (data part), the CST\_ROIs are the images which are generated by manual delineation by the ITK-SNAP. The number of the tracks set as 1000, the radius is the 1.5, the cortex label of left precentral gyrus is 49 and the right precentral gyrus is 50. Besides, the left labels of the three ROIs are 1, 3 and 5, and the right labels are the 2, 4, and 6. The left and the right CST fibers are computed separately, the cortex label and the ROIs labels should be consistent with the above parameters. 
\begin{figure*}
\begin{subfigure}{0.30\textwidth}
\includegraphics[width=\linewidth]{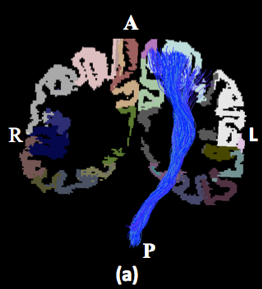}
\end{subfigure}
\hspace*{\fill} 
\begin{subfigure}{0.31\textwidth}
\includegraphics[width=\linewidth]{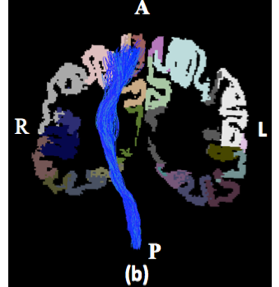}
\end{subfigure}
\hspace*{\fill} 
\begin{subfigure}{0.31\textwidth}
\includegraphics[width=\linewidth]{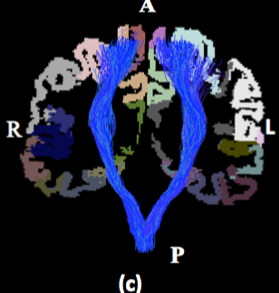}
\end{subfigure}
\caption{The coronal slice of the reconstructed CST fibers. It overlays the fibers with the cortex label image to show the projection of the fiber tracts into the precentral gyrus. (a): the left CST fibers, (b): the right CST fibers, (c): both the left and right CST fibers (A: anterior, P: Posterior, R: right, L: left).  } \label{fig:3}
\end{figure*}

\subsection{Outlier tracts remove}
   The result of the workflow is a tck file, we convert it into the trk file by Matlab since the BrainSuite software \cite{shattuck2002brainsuite} cannot recognize the tck files. In addition, some fibers are not normal and may be too long and project to other areas. Firstly, we filter the outlier tracts, and then some obvious abnormal tracts are removed using the BrainSuite 16 a1 track filtering options.

\begin{figure}
\begin{subfigure}{0.2\textwidth}
\includegraphics[width=\linewidth]{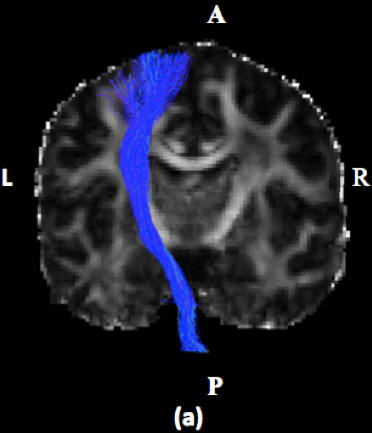}
\end{subfigure}
\hspace*{\fill} 
\begin{subfigure}{0.2\textwidth}
\includegraphics[width=\linewidth]{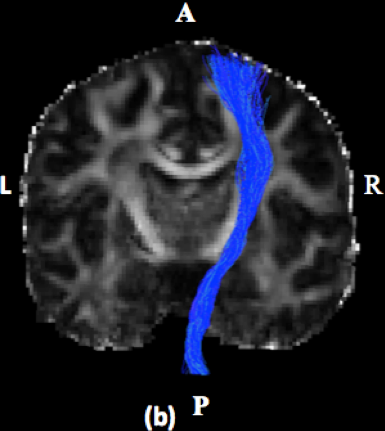}
\end{subfigure}
\hspace*{\fill} 
\caption{The coronal slice of the reconstructed CST fibers. It overlays the fibers with the FA image to show the projection of the fiber tracts into the precentral gyrus. (a): the left CST fibers, (b): the right CST fibers (A: anterior, P: Posterior, R: right, L: left).   } \label{fig:4}
\end{figure}

\section{Results}
   The reconstructed CST fibers are visualized by the BrainSuite 16 a1, to show the anatomical validity of the reconstruction results, we overlay the CST fiber bundles with the FA image. Fig. 4 and 5 show the projection to the precentral gyrus by overlaying the fiber tracts with the cortex label and FA images. Fig. 6 shows the results of the CST fibers through the posterior of the internal capsule and the thalamus. Fig. 7 shows the results of the cerebral peduncle by overlaying the fiber bundles with the FA image. And the Fig. 8 overlays the fiber bundle with the FA image to the anterior pontine area. These results demonstrate that the fiber bundles are anatomically correct. Furthermore, these results prove that the FOD-based tractography can provide a smooth, clear and reliable reconstruction of the CST fibers.

\begin{figure}
\begin{subfigure}{0.2\textwidth}
\includegraphics[width=\linewidth]{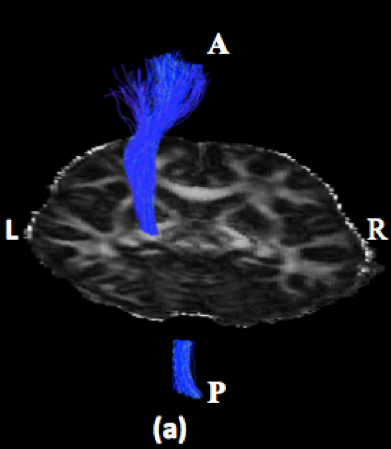}
\end{subfigure}
\hspace*{\fill} 
\begin{subfigure}{0.2\textwidth}
\includegraphics[width=\linewidth]{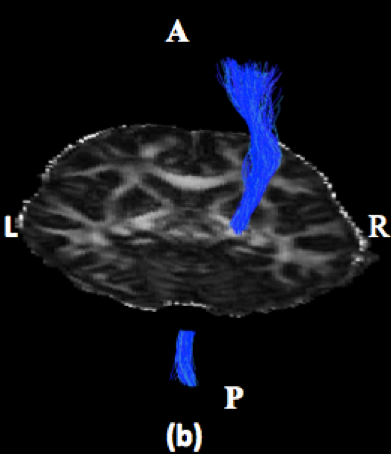}
\end{subfigure}
\hspace*{\fill} 
\caption{The axial slice of the reconstructed CST fibers. It overlays the fiber bundle with the FA image that shows the internal capsule and the thalamus. (a): the left CST fibers, (b): the right CST fibers (A: anterior, P: Posterior, R: right, L: left).     } \label{fig:5}
\end{figure}

\begin{figure}
\begin{subfigure}{0.2\textwidth}
\includegraphics[width=\linewidth]{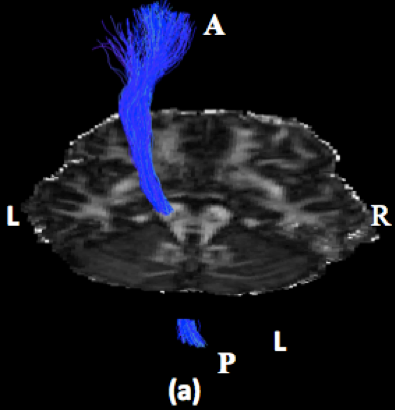}
\end{subfigure}
\hspace*{\fill} 
\begin{subfigure}{0.2\textwidth}
\includegraphics[width=\linewidth]{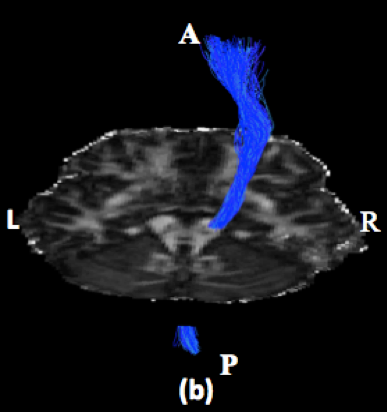}
\end{subfigure}
\hspace*{\fill} 
\caption{The axial slice of the reconstructed CST fibers. It overlays the fibers with the FA image to show the middle brain (mainly the cerebral peduncle). (a): the left CST fibers, (b): the right CST fibers (A: anterior, P: Posterior, R: right, L: left).   } \label{fig:7}
\end{figure}

\begin{figure}
\begin{subfigure}{0.18\textwidth}
\includegraphics[width=\linewidth]{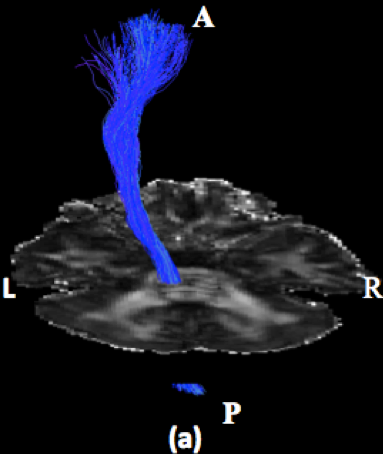}
\end{subfigure}
\hspace*{\fill} 
\begin{subfigure}{0.2\textwidth}
\includegraphics[width=\linewidth]{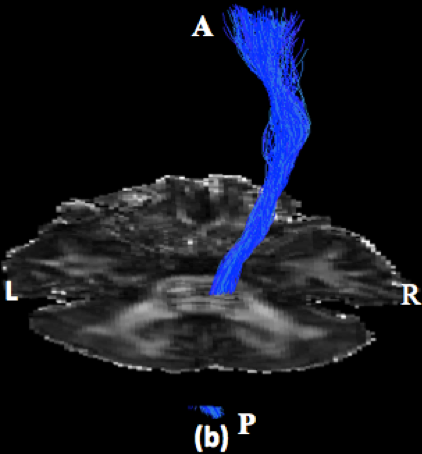}
\end{subfigure}
\hspace*{\fill} 
\caption{The axial slice of the reconstructed CST fibers. It overlays the fibers with the FA image to show the anterior pontine area. (a): the left CST fibers, (b): the right CST fibers (A: anterior, P: Posterior, R: right, L: left).    } \label{fig:6}
\end{figure}

\section{Discussion}
Comparing our results with \cite{kristo2013reliability,huttlova2014abnormalities}, our reconstruction results show more fibers, which shows the high accuracy of our methods. One limitation of the paper is that we only choose three ROIs (the posterior limb of the internal capsule, the cerebral peduncle, and the anterior pontine area); there are more regions which can be explored, such as top of the lateral ventricle (CSoLV) though it might not be as important as above three ROIs, it still can be an interesting direction. Our study also had a limited sample size. However, Our tractography findings suggest the FODs reconstruction method can accurately reconstruct the CST fibers; thus, the power of our study is sufficient.
For future work, we should explore our reconstruction of the CST in recovery from the SCI injury.

\section{Conclusion}
In this paper, we reconstruct the Corticospinal Tract using fiber orientation distributions tractography. Comparing with determined orientation tensor-based tractography, the main direction of FODs is based on  probabilistic direction. We  manually  delineate  the  three  ROIs  (the  posterior  limb  of the  internal  capsule,  the  cerebral  peduncle,  and  the  anterior pontine area) by the ITK-SNAP software and use the pipeline software reconstructs both the left and right sides of the CST fibers. Our results demonstrate that FOD-based tractography can show correct anatomical CST fiber bundles. The results of our study show that the accurate reconstruction of CST can be applied in improving both diagnostics and treatment of Spinal cord injury in the future.


\section*{Acknowledgment}

We want to thank professor Yonggang Shi for guiding  us  to  do  research  on  this  topic  and  give  us  valuable suggestions in explaining the topic more professionally.



%


\bibliographystyle{unsrt}
\bibliography{references}

\end{document}